# Improved diffusing wave spectroscopy based on the automatized determination of the optical transport and absorption mean free path

Chi Zhang[1], Mathias Reufer[2,†], Danila Gaudino[2] and Frank Scheffold[1,*]

[1]*Department of Physics, University of Fribourg, Fribourg 1700, Switzerland*
[2] *LS Instruments AG, Fribourg 1700, Switzerland*

Diffusing wave spectroscopy (DWS) can be employed as an optical rheology tool with numerous applications for studying the structure, dynamics and linear viscoelastic properties of complex fluids, foams, glasses and gels. To carry out DWS measurements, one first needs to quantify the static optical properties of the sample under investigation, *i.e.* the transport mean free path $l^*$ and the absorption length $l_a$. In the absence of absorption this can be done by comparing the diffuse optical transmission to a calibration sample whose $l^*$ is known. Performing this comparison however is cumbersome, time consuming and prone to mistakes by the operator. Moreover, already weak absorption can lead to significant errors. In this paper, we demonstrate the implementation of an automatized approach, based on which the DWS measurement procedure can be simplified significantly. By comparison with a comprehensive set of calibration measurements we cover the entire parameter space relating measured count rates ($CR_t$, $CR_b$) to ($l^*$, $l_a$). Based on this approach we can determine $l^*$ and $l_a$ of an unknown sample accurately thus making the additional measurement of a calibration sample obsolete. We illustrate the use of this approach by monitoring the coarsening of a commercially available shaving foam with DWS.

---

[*]Corresponding author: Frank.Scheffold@unifr.ch
[†]Co-corresponding author; mathias.reufer@lsinstruments.ch

# 1. Introduction

Diffusing wave spectroscopy (DWS) is a modern optical technique derived from dynamic light scattering (DLS), allowing the measurements of thermally driven dynamics in strongly scattering media (Maret and Wolf, 1987; Pine *et al.*, 1988; Pine *et al.*, 1990). One of the most significant advantages of this technique is that it allows the measurement of the dynamics over a large range of time scales ($10^{-7}$ to 10 s) and at much shorter length scales (down to ~ 1 nm) than traditional single light scattering experiments (Scheffold and Schurtenberger, 2003; Weitz and Pine, 1993; Zhu *et al.*, 1992). Moreover, DWS can be used to perform non-invasive microrheology, extracting rheological properties on the micron scale without making actual contact to the sample (Mason and Weitz, 1995; Waigh, 2016; Furst and Squires, 2017). Since it was developed in the late 1980's, DWS has been widely used in the study of soft matter, such as colloidal suspensions, microgels, emulsions, foams, and biological media (Li *et al.*, 2005; Palmer *et al.*, 1999; Scheffold, 2002; Scheffold *et al.*, 2010; Lee *et al*, 2013).

Since the samples studied with DWS are turbid, one needs to calculate the intensity autocorrelation function of the multiple scattered light. This can be done by considering the diffuse propagation of photons along paths of different lengths $s$. Each path length $s$ contributes to the intensity correlation function in a defined way and its weight is determined by the path length distribution. It has been shown that, for a given cuvette size and geometry, the path length distribution only depends on the transport mean free path $l^*$ of the sample and the absorption length $l_a$ (Pine *et al.*, 1988). Using the diffusion equation to model the transport of light in an opaque medium, analytic expressions for the DWS intensity correlation function in dependence of ($l^*$, $l_a$) and of the thermally driven colloidal dynamics can be obtained. For the case of bead motion in a viscous or viscoelastic matrix the internal dynamics is given by the particle mean square displacement (MSD) $\langle\Delta\vec{r}^2(t)\rangle$ (Weitz and Pine, 1993). Therefore, to extract the dynamic properties of the sample, one first needs to quantify the static, ensemble averaged optical properties ($l^*$, $l_a$) of the sample. Since the dynamics of the target sample is not known, this is normally done by comparing the optical properties of the sample with another sample whose $l^*$ is either known, can be calculated or most often can be measured. For example, with the commercial DWS-RheoLab instrument (LS Instruments, Switzerland), an extra sample denoted the 'calibration standard' has to be measured prior to the actual measurement of interest. This calibration standard must have similar turbidity (*i.e.* similar $l^*$) as the target sample and its dynamics must be known. Typically, for monodisperse particles with known diameter dispersed in water $\langle\Delta\vec{r}^2(t)\rangle = 6D_B t$ where $D_B$ is the known Brownian diffusion coefficient of the particles. Currently, the instrument first measures $l^*$ of the calibration standard by fitting the correlation function. Then comparing the transmission count rate of the two samples, the instrument then can determine $l^*$ for the target sample, with an accuracy of approximately 5-10% [Technical note, LS Instruments].

However, in some cases, the target sample could be absorbing at the incident laser wavelength. For this case, additional knowledge about the sample absorption length $l_a$ is required to extract *e.g.* the MSD. Until now it was not possible to determine the absorption length $l_a$ with the commercial instrument and therefore the rheological properties of absorbing, or colored, samples could not be accurately characterized.

In this paper, we implement an automatized calibration of the DWS measurement procedure. With this approach, we can determine not only the transport mean free path $l^*$ directly but also the absorption length $l_a$ of the sample, by simply measuring the photon count rate of both the light scattered in transmission and backscattering geometry. In this way, the calibration step as well as the preparation of the calibration standard will be no longer needed. This allows us to perform DWS experiments more conveniently and extend the range of application of quantitative DWS and DWS microrheology to absorbing samples, as long as $l_a \gg l^*$.

# 2. Methods

The photon count rates of light in the transmission $CR_t$ and backscattering geometry $CR_b$ are easily measurable. Here we denote with $CR_b = CR_{b,VH} + CR_{b,VV}$ the sum of the count rates measured independently for the polarization preserving (VV) and the cross polarized (VH) detection channel. Therefore, if we can derive an unambiguous mathematical relation between ($CR_t$, $CR_b$) and ($l^*$, $l_a$), we will be able to quantify $l^*$ and $l_a$ by simply measuring the photon count rates. Numerically we expect this approach to work even if this relationship is very complicated for a specific experimental configuration (finite cuvette width, residual surface reflectivity and so on).

We begin by summarizing again the most important concepts governing light transport and intensity fluctuations and correlations in the frame of DWS which can are obtained by summing over the contribution from all photons arriving at the detector via different optical paths. Denoting $P(s)$ as the probability that the photon follows the path $s$ in the absence of absorption, we can write the field correlation function (with absorption) as (Weitz and Pine, 1993)

$$g_1(t) = \frac{1}{C}\int_0^\infty P(s)e^{-s/l_a}e^{-2(t/\tau)(s/l^*)}ds \quad (1)$$

where $\tau = 1/k_0^2 D_0$ is the characteristic decay time and $C$ is the normalizing constant. For simplicity, we consider Brownian motion but the same equations apply more generally to other dynamic processes such as bubble rearrangements in foams or tracer bead motion in a viscoelastic medium, for the latter $t/\tau$ is simply replaced by $k_0^2 \langle \Delta \vec{r}^2(t)\rangle/6$, where $k_0 = 2\pi n/\lambda$ (laser wavelength $\lambda$) denotes the wavenumber in the scattering medium with a refractive index n. The field correlation function $g_1(t)$ is obtained from the measured intensity correlation function $g_2(t)$ in the common way using the Siegert relation $g_2(t) = 1 + \beta g_1^2(t)$. The instrument specific coherence factor $\beta \sim 0.9$ is obtained experimentally by extrapolation $t \to 0$. The first exponential term in Eq. (1) describes the photon loss due to possible absorption, and the second exponential sums up the decay of correlation due to multiple interactions with scattering particles. Both terms are weighted by the probability that the light follows a path of length $s$. It has been shown that if the incident laser beam is expanded to fill the full surface of the sample (flat cell geometry, thickness L), we can write the correlation function in transmission as (Weitz and Pine, 1993)

$$g_1(t) = \frac{1}{C_t}\frac{\frac{L/l^*+4/3}{z_0/l^*+2/3}[sinh(\frac{z_0}{l^*}\Delta) + \frac{2}{3}\Delta cosh(\frac{z_0}{l^*}\Delta)]}{(1+\frac{4}{9}\Delta^2)sin(\frac{L}{l^*}\Delta) + \frac{4}{3}\Delta cosh(\frac{L}{l^*}\Delta)} \quad (2)$$

and for the backscattering direction as

$$g_1(t) = \frac{1}{C_b}\frac{sinh[\Delta(\frac{L}{l^*} - \frac{z_0}{l^*})] + \frac{2}{3}\Delta cosh[\Delta(\frac{L}{l^*} - \frac{z_0}{l^*})]}{(1+\frac{4}{9}\Delta^2)sinh(\frac{L}{l^*}\Delta) + \frac{4}{3}\Delta cosh(\frac{L}{l^*}\Delta)} \quad (3)$$

where $C_t$, $C_b$ are the normalizing constants which ensure $g_1(t=0) = 1$, and $z_0$ describes the distance inside the sample where the light propagation can be considered as diffusive. The latter can be roughly estimated as $z_0 \approx l^*$. In these equations, $\Delta = (6t/\tau + 3l^*/l_a)^{1/2}$ denotes the sum of the particle displacements and the absorption contribution. The normalizing constants $C_t$, $C_b$ are related to the transmission and backscattering count rate as discussed later in the text. For a ideal, non-absorbing sample, i.e. $l_a \to \infty$, always $C_t, C_b = 1$.

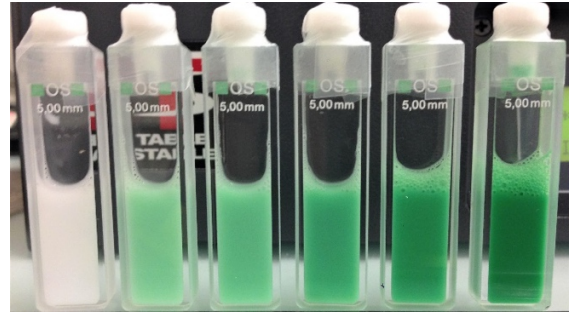

**Fig.1.** Polystyrene particles, diameter 910 nm, in cuvettes with inner thickness $L$ = 5 mm, width 9.5mm (100-OS, Hellma, Germany). The concentration of green dye, from left to right, is 0, 5 µl/ml, 10 µl/ml, 20 µl/ml, 40 µl/ml and 80 µl/ml, corresponding to an absorption length of infinity, 146 mm, 73 mm, 36.5 mm, 18.8 mm and 9.4 mm, respectively. The transport mean free path is 0.55mm in all cases.

Next, we treat the transmission case explicitly. The backscattering case can be derived in the same way. Since DWS only can be applied in the diffusive limit $L > l^*$, the measured transmission count rate recorded at some scattering angle, not too far from the normal, is proportional to the total transmission $T$ (total transmitted flux). We denote the transmission rate of an ideal non-absorbing sample as $T_{ideal}$ ($l^*$). When some absorption is present (quantified by the absorption length $l_a$), the transmission coefficient of this sample then can be written as

$$T(l^*, l_a) = T_{ideal}(l^*)\int_0^\infty P(s)e^{-s/l_a}ds \quad (4)$$

Comparing Eq. (4) with Eq. (1) in the limit of $t = 0$, we can easily see that

$$T(l^*, l_a) = T_{ideal}C_t(l^*, l_a) \quad (5)$$

, where the normalizing constant $C_t$ ($l^*$, $l_a$) can be expressed as a function of $l^*$ and $l_a$ using Eq. (2) (for t=0). It is worth pointing out that here '*ideal*' stands for the asymptotic case of a nonabsorbing sample contained in an infinitely wide cuvette and illuminated by an incident plane wave. However, in reality, the width of the cuvettes is finite. Here, we use cuvettes of internal width 9.5mm shown in Fig. 1. Thus, some loss of the photons on both sides of the cuvette is inevitable. Such photon loss can be treated equal to being *absorbed*. To characterize such photon loss, we introduce a new parameter $l_{a0}$ as the 'photon loss length' of the cuvette, which is an inherent property of the cuvette and is found to be independent of the sample contained within. Using a nonabsorbing sample whose $l^*$ is known, $l_{a0}$ can be obtained by fitting the measured correlation function with Eq. (2), with $l_{a0}$ as the only adjustable parameter. We found in our experiments that measurements of nonabsorbing samples with different $l^*$ gave similar $l_{a0}$ when the same cuvette thickness $L$ was used, which confirms that treating the photon loss due to the cuvette geometry as suggested is feasible. The values for $l_{a0}$ we find are of the order of 30-40mm, independent of $l^*$ and only weakly dependent on $L$. We note that this empirical observation is probably related to the recent observation that the mean path length in multiple light scattering of waves is independent of $l^*$ (Pierrat et al., 2014). While beyond the scope of this article, we plan to address this interesting question in more detail in future work.

Now, let us consider the case of a non-absorbing sample with known $l^*$. The count rate of both $CR_t$ and $CR_b$ can be easily measured with DWS. By measuring the $CR$ for samples with different $l^*$, an empirical relation between the $CR$ and $l^*$ can be established. We denote as $CR_{t0}(l^*, l_{a0})$ and $CR_{b0}(l^*, l_{a0})$ the $l^*$ dependent count rate in a cuvette with a finite $l_{a0}$ obtained independently.

After these general remarks, we now turn our attention to the implementation of the automatized calibration procedure. Consider $CR_t$ and $CR_b$ as the transmission and backscattering count rate measured by DWS for the target sample $(l^*, l_a)$. Substituting $CR_{t0}(l^*, l_{a0})$ and $CR_{b0}(l^*, l_{a0})$ into Eq. (5), we can obtain the following equations

$$\frac{CR_t}{CR_{t0}(l^*, l_{a0})} = \frac{T(l^*, l'_a)}{T(l^*, l_{a0})} = \frac{C_t(l^*, l'_a)}{C_t(l^*, l_{a0})}, \quad (6)$$

$$\frac{CR_b}{CR_{b0}(l^*, l_{a0})} = \frac{R(l^*, l'_a)}{R(l^*, l_{a0})} = \frac{C_b(l^*, l'_a)}{C_b(l^*, l_{a0})} \quad (7)$$

where $R$ is the backscattering rate and $l_a'$ ($1/l_a' = 1/l_{a0} + 1/l_a$) is the total photon absorption (or loss) of the target sample. $C_t$ and $C_b$ can be easily expressed as functions of $l^*$ and $l_a$ with Eqs. (2) and (3), for $t = 0$. $CR_{t0}(l^*, l_{a0})$ and $CR_{b0}(l^*, l_{a0})$ also can be expressed as function of $l^*$ and $l_a$ empirically, as discussed above (details in the next section). Therefore, $l^*$ and $l_a$ can be obtained by solving Eqs. (6) and (7) numerically.

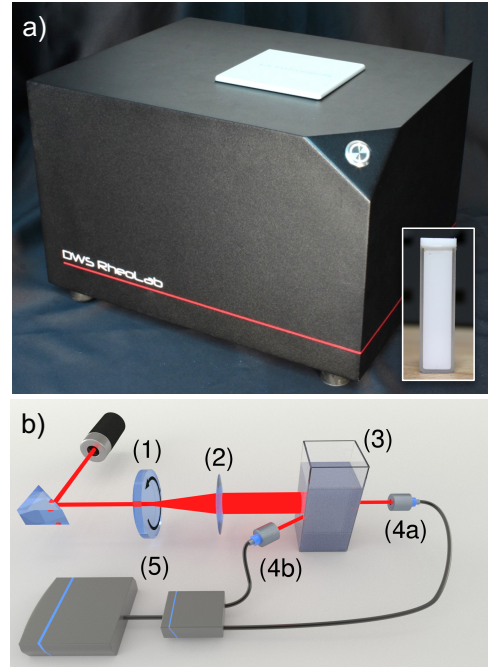

**Fig. 2.** Diffusing wave spectroscopy setup. (a) Commercial DWS instrument (DWS RheoLab III, LSInstruments, Switzerland). The inset shows the foam sample contained in a L=5mm cuvette before loading the cuvette in the instrument. (b) Optical configuration: A coherent light source (Laser, Cobolt, Sweden, $\lambda$ = 687 nm) is directed to the surface of a ground glass diffuser (2) mounted on a stepper motor. The speckle beam created by the diffuser is used to illuminate a sample cuvette containing the sample of interest (3). Single mode fiber receivers (4a and 4b) collect the scattered light either in transmission or backscattering and direct it to a single photon counting module and digital correlator (5).

## 3. Experiments and result

To determine the empirical relations $CR_{t0}(l^*, l_{a0})$ and $CR_{b0}(l^*, l_{a0})$ and to verify the automatized calibration approach, we perform a set of DWS-Experiments ($\lambda$ = 687 nm) on suspensions of polystyrene latex spheres (PS) using a commercial DWS instrument (DWS RheoLab III, LS Instruments, Switzerland), shown in Fig. 2. Monodisperse PS particles are dispersed in 2 mM water based SDS (sodium dodecyl sulfate) solution to prevent aggregation. In our experiments, particles with diameter 910 nm and 190 nm are both studied to exclude any possible influence of the particle size. Absorption is introduced by adding different amounts of green food dye. We checked that this food dye does not adsorb to the PS particle surface. The absorption of the dye was characterized with a simple transmission intensity measurement of the dye solvent (no particles). The transmission intensity of such solvent can be well described by the Beer-Lambert law:

$$-log(\frac{I_t}{I_0}) = k \cdot c \cdot L \tag{8}$$

where $I_0$ is the transmission intensity using pure water with the same cuvette thickness $L$, $k$ is the absorption coefficient and $c$ is the concentration of the dye (µl/ml). By measuring the transmission intensity at the same wavelength as for the DWS experiment using a UV-VIS spectrometer, we found the absorption coefficient of the dye is k = 1.37×10$^{-3}$ ml/(mm·µl). Therefore, the absorption length of a sample can be calculated as $l_a = 1/(k \cdot c)$.

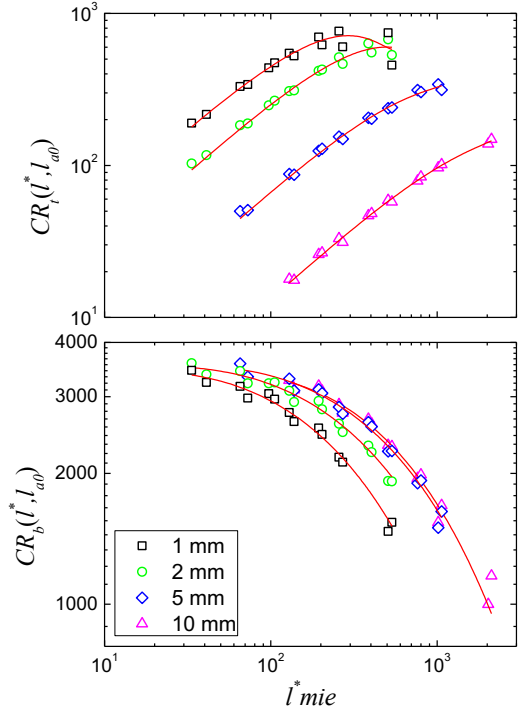

**Fig.3.** Count rate of non-absorbing sample measured in transmission (upper) and backscattering (lower) with particle size of both 910 nm and 190 nm as a function of transport calculated transport mean free path (in µm). Each cuvette thickness L is presented with a different color as shown in the legend. Symbols denote the experimental data and solid lines show the best fit to the data based on Eqs. (9) and (10).

For establishing the empirical relation of $CR_{t0}$ ($l^*$, $l_{a0}$) and $CR_{b0}$ ($l^*$, $l_{a0}$), for a given cuvette thickness L, we first used non-absorbing samples with both 910 nm and 190 nm particles. For each cuvette thickness, seven samples with different volume fractions (for each particle size) were measured, covering $L/l^*$ from less than 5 to more than 80. Since the volume fraction of particles is known, we can calculate $l^*$ of the sample using Mie scattering theory as implemented in ref. (Ochoa, 2004), in the following denoted as $l^*_{mie}$. As shown in Fig. 3, for the same cuvette thickness, the count rates for particles with the same ratio $L/l^*$, but different size, are basically indistinguishable (hence they are presented using the same symbols), except for cases when $L/l^*$ is smaller than five. Only for the L = 1 mm data (black open squares) in the upper panel, one can see that the count rate starts to divide into two branches (upper: 910 nm; lower: 190 nm) when $l^*_{mie}$ is around 200 µm. We did not further explore these differences since accurate DWS measurements are normally restricted to samples with $L/l^* > 5$ (Kaplan et al.1993). For $L/l^* < 5$ deviations from the diffusion approximation are expected but a detailed analysis of this regime is beyond the scope of our work. The measurements however provide a good estimate for the lower limit $L/l^* > 5$ where our approach can be applied safely.

Next, we establish the empirical relation $CR_{t0}$ ($l^*$, $l_{a0}$) and $CR_{b0}$ ($l^*$, $l_{a0}$) for each cuvette thickness $L$. To this end, we fit the measured data with the following empirical equations.

$$CR_{t0}(l^*, l_{a0}) = A_t(l^* + B_t l^{*2} + D_t l^{*3}), \tag{9}$$

$$CR_{b0}(l^*, l_{a0}) = \frac{A_b}{l^{*D_b}} e^{-1.9\sqrt{3l^*/B_b}} \tag{10}$$

where $A$, $B$, and $D$ are the fitting parameters. As shown in Fig. 3, Eqs. (9) and (10) describes the data well.

We are now ready to apply the automatized calibration, since $CR_{t0}$ ($l^*$, $l_{a0}$) and $CR_{b0}$ ($l^*$, $l_{a0}$) as function of ($l^*$, $l_{a0}$) are known. We test the accuracy of our approach with samples whose $l^*$ and $l_a$ are known. To this end, we prepare our sample with the green dye solution instead of pure water, Fig. 1. For each cuvette thickness, at least 4 volume fractions are measured with at least 4 well controlled dye concentrations, covering $l_a$ from about 5 mm to more than 200 mm. By measuring the count rate in transmission and backscattering, we quantify $l^*$ and $l_a$, following the approach just described. Here we compare the experimental values to the known reference values: $l^*$ from the Mie calculator, or $l_a$ calculated from the known dye concentration using $l_a = 1/(k \cdot c)$. As shown in Fig. 4, the experimentally determined $l^*$ and $l_a$ show no systematic dependence on neither cuvette thickness $L$ nor $L/l^*$ over the region of interest. Fitting the corresponding probability distribution to a Gaussian distribution, the accuracy with respect to determining $l^*$ and $l_a$ is approximately 8% and 14%, respectively, in the region of $5 < L/l^* < 100$. Compared to the previous calibration method, which only can be applied to non-absorbing samples, the auto-calibration approach has similar accuracy with respect to determining $l^*$. Moreover, our approach can characterize, at the same time, the absorption length $l_a$, which was not possible with the conventional calibration method.

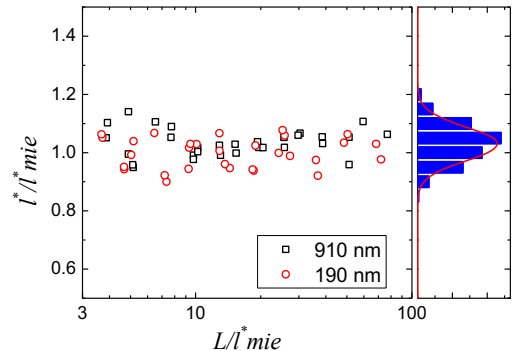

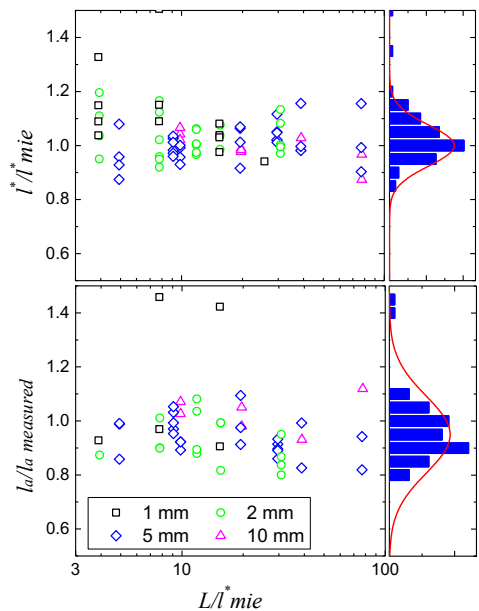

**Fig. 4.** $l^*$ (upper) and $l_a$ (lower) characterized with the auto-calibration approach compared with the known value. Each cuvette thickness L is presented with different colors as shown in the legend. The probability distributions are shown with corresponding histogram on the right side of the graphs. Experimental measures are presented with blue bars and the fittings to a Gaussian distribution are shown with solid lines.

Finally, we verified again, using only non-absorbing samples, whether there exists a direct influence of the particle size on our results that may lead to a hidden systematic error. We compare the experimentally determined values with $l^*_{mie}$, now independently for the d = 910 and d = 190 nm samples. As shown in Fig. 5, $l^*$ shows no systematic dependence on the particle size in the region of interest. Fitting its probability distribution to a Gaussian distribution, the accuracy is about 5% in the region of $5 < L/l^* < 100$, slightly better than to the whole set of data including absorbing samples.

**Fig. 5.** $l^*$ of the non-absorbing sample characterized with the auto-calibration approach compared with the known value. The data of 910 nm diameter particles is shown in black and of 190 nm particles in red. The probability distribution is shown with a histogram on the right side of the graphs. Experimental measures are presented with blue bars and its fitting to a Gaussian distribution is shown with a solid line.

## 4. Application example: Coarsening of a commercial shaving foam

To illustrate the use of our approach we have applied the new routine to monitor the coarsening of a commercially available shaving foam (Gillette, USA) with DWS. The experimental device, optical configuration and the foam sample are shown in Fig. 2. Foams are dense assemblies of gas bubbles separated by reflecting liquid thin films that meet at the connections, known as the plateau borders. Due to the familiar coarsening of the bubbles due to gas exchange across the film interfaces and subsequent bubble rearrangements we expect both an evolution in the relaxation time $\tau$ and an increase of $l^*$ (Durian et al., 1991a). If one assumes that the size distribution of the foam bubbles remains the same, one can relate directly the transport mean free path $l^*$ to the mean bubble radius $\langle R \rangle$. The proportionality $l^* \propto \langle R \rangle$ can be derived from the self-similar coarsening of the foam.

The measurements are done by loading a standard L = 5 mm thickness glass cuvette, width 9.5 mm, with Gillette foam. The DWS RheoLab instrument is operated in dual multi-tau/echo mode with a total measurement time of approximately 300 seconds for each run. The time interval between two runs is about 630 s. In every run, the RheoLab first characterize $l^*$ and $l_a$ based on the experimentally determined count rates $CR_b$, $CR_t$ and subsequently the correlation function $g_1(t)$ in transmission is recorded. In contrast to earlier pioneering DWS experiments on similar foams, ref. (Durian et al., 1991a), using the new automatized approach, there is no need to repeat the measurement for several different L to extract $l^*$ and verify that $l_a \gg L$. This notably reduces the measurement time for the present example and similar time savings are expected for all slowly evolving samples where both the internal dynamics and $(l^*, l_a)$ evolve with time. In addition, the new algorithm simplifies the experiment and leads to more accurate results. This is particularly useful for DWS users not deeply familiar with the optics of diffuse light scattering experimentation, since they can now determine the key parameters $l^*$ and $l_a$ with ease.

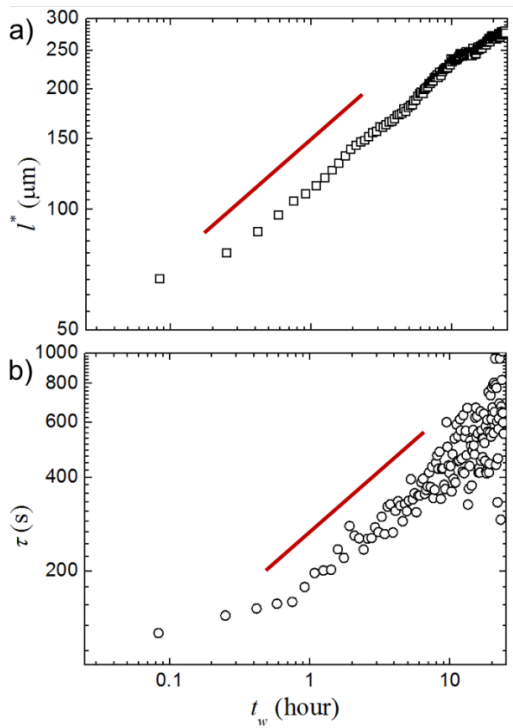

**Fig. 6.** Time evolution of the transport mean free path $l^*$ and the relaxation time $\tau$ of a Gillette shaving foam, contained in a L = 5mm cuvette, continuously monitored by DWS over more than 10 h. The absorption length extracted using the new routine is found to be nearly infinite ($l_a \gg 10L$, i.e. beyond the detection limit) indicating the absence of absorption within the sample as expected given the familiar white appearance of shaving foam. As noted earlier the optical transmission $T \propto l^*/L$ and thus $l^* \propto \langle R \rangle$, the mean bubble radius, scales as a power law. Equally the characteristic relaxation time $\tau$ increases as a power law for times larger than about 20 min. For more details see refs. (Cohen-Addad and Höhler, 2001; Durian et al., 1991a; 1991b; Sessoms et al., 2010).

As reported previously we find that $g_1(t)$ decays nearly exponentially, indicating a single relaxation process. We obtain the relaxation time $\tau$ from a fit of Eq. (1) to $g_1(t)$. The results both for $l^*$ and $\tau$ as a function of the waiting time $t_w$ are shown in Fig. 6. The results obtained are in overall agreement with previous, more detailed, DWS studies of the foam coarsening process (Cohen-Addad and Höhler, 2001; Durian et al., 1991a; 1991b; Sessoms et al., 2010). This example, taken from a single run of the instrument, demonstrates that DWS can now be an even more convenient tool to monitor slowly evolving complex fluids.

## 5. Conclusions

In this paper, we present an automatized calibration approach to diffusing wave spectroscopy (DWS) measurements. The successfully implemented auto-calibration approach improves the accuracy and simplifies the measurement procedure significantly. With this approach, we are now able to determine the transport mean free path $l^*$ and the absorption length $l_a$ of the sample, by simply measuring the count rate of the transmission and backscattering in a commercial instrument. In this way, previously required additional, time consuming calibration steps, which usually also required careful preparation of a calibration and/or cuvettes with different path lengths L, are now obsolete. Our control experiments show that this auto-calibration approach is practical and accurate. It is robust against scattering from particles of different size and can work over a wide range of optical densities $L/l^*$. We demonstrate that the automatized calibration provides an accuracy of around 8% and 5% on determining $l^*$, for absorbing and non-absorbing samples, respectively. Moreover, it allows us to also determine the absorption length $l_a$, which previously was not possible, with an accuracy of about 14%.

## Acknowledgements

This project was financially supported by the CTI (Innovation promotion agency of the Swiss Confederation) under grant 19278.1 PFNM-NM.